\documentclass[a4paper]{article}

\usepackage{icrc2013}

\title{KASCADE-Grande measurements of energy spectra for elemental groups of cosmic rays}

\newcommand{\etal}{\MakeLowercase{\textit{et al. }}} 
\shorttitle{D. Fuhrmann \etal KASCADE-Grande measurements of energy spectra}

\authors{
D.~Fuhrmann$^{8,c}$,
W.D.~Apel$^{1}$,
J.C.~Arteaga-Vel\'azquez$^{2}$,
K.~Bekk$^{1}$,
M.~Bertaina$^{3}$,
J.~Bl\"umer$^{1,4}$,
H.~Bozdog$^{1}$,
I.M.~Brancus$^{5}$,
E.~Cantoni$^{3,6,a}$,
A.~Chiavassa$^{3}$,
F.~Cossavella$^{4,b}$,
C.~Curcio$^{3}$,
K.~Daumiller$^{1}$,
V.~de Souza$^{7}$,
F.~Di~Pierro$^{3}$,
P.~Doll$^{1}$,
R.~Engel$^{1}$,
J.~Engler$^{1}$,
B.~Fuchs$^{4}$,
H.J.~Gils$^{1}$,
R.~Glasstetter$^{8}$,
C.~Grupen$^{9}$,
A.~Haungs$^{1}$,
D.~Heck$^{1}$,
J.R.~H\"orandel$^{10}$,
D.~Huber$^{4}$,
T.~Huege$^{1}$,
K.-H.~Kampert$^{8}$,
D.~Kang$^{4}$, 
H.O.~Klages$^{1}$,
K.~Link$^{4}$, 
P.~{\L}uczak$^{11}$,
M.~Ludwig$^{4}$,
H.J.~Mathes$^{1}$,
H.J.~Mayer$^{1}$,
M.~Melissas$^{4}$,
J.~Milke$^{1}$,
B.~Mitrica$^{5}$,
C.~Morello$^{6}$,
J.~Oehlschl\"ager$^{1}$,
S.~Ostapchenko$^{1,d}$,
N.~Palmieri$^{4}$,
M.~Petcu$^{5}$,
T.~Pierog$^{1}$,
H.~Rebel$^{1}$,
M.~Roth$^{1}$,
H.~Schieler$^{1}$,
S.~Schoo$^{1}$,
F.G.~Schr\"oder$^{1}$,
O.~Sima$^{12}$,
G.~Toma$^{5}$,
G.C.~Trinchero$^{6}$,
H.~Ulrich$^{1}$,
A.~Weindl$^{1}$,
J.~Wochele$^{1}$,
J.~Zabierowski$^{11}$ \\
KASCADE-Grande Collaboration
}

\afiliations{
$^1$ Institut f\"ur Kernphysik, KIT - Karlsruher Institut f\"ur Technologie, Germany\\
$^2$ Universidad Michoacana, Instituto de F\'{\i}sica y Matem\'aticas, Morelia, Mexico\\
$^3$ Dipartimento di Fisica, Universit\`a degli Studi di Torino, Italy\\
$^4$ Institut f\"ur Experimentelle Kernphysik, KIT - Karlsruher Institut f\"ur Technologie, Germany\\
$^5$ National Institute of Physics and Nuclear Engineering, Bucharest, Romania\\
$^6$ Osservatorio Astrofisico di Torino, INAF Torino, Italy\\
$^7$ Universidade S$\tilde{a}$o Paulo, Instituto de F\'{\i}sica de S\~ao Carlos, Brasil\\
$^8$ Fachbereich Physik, Universit\"at Wuppertal, Germany\\
$^9$ Department of Physics, Siegen University, Germany\\
$^{10}$ Dept. of Astrophysics, Radboud University Nijmegen, The Netherlands\\
$^{11}$ National Centre for Nuclear Research, Department of Cosmic Ray Physics, Lodz, Poland\\
$^{12}$ Department of Physics, University of Bucharest, Bucharest, Romania\\
\scriptsize{
$^{a}$ now at: Istituto Nazionale di Ricerca Metrologia, INRIM, Torino; \\
$^{b}$ now at: Max-Planck-Institut Physik, M\"unchen, Germany; \\
$^{c}$ now at: University of Duisburg-Essen, Duisburg, Germany;  \\
$^{d}$ now at: University of Trondheim, Norway.
}
}

\email{fuhrmann@uni-wuppertal.de} 

\abstract{The KASCADE-Grande experiment, located at KIT-Karlsruhe, Germany, consists of a large scintillator array for measurements of charged particles, $N_{ch}$, and of an array of shielded scintillation counters used for muon counting, $N_{\mu}$. KASCADE-Grande is optimized for cosmic ray measurements in the energy range $10^{16}$~eV to $10^{18}$~eV, thereby enabling the verification of a knee in the iron spectrum expected at approximately $10^{17}$~eV. Exploring the composition in this energy range is of fundamental importance for understanding the transition from galactic to extragalactic cosmic rays.\\
Following earlier studies of elemental spectra reconstructed in the knee energy range from KASCADE data, we have now extended these measurements to beyond $10^{17}$~eV. By analysing the two-dimensional shower size spectrum $N_{ch}$ vs. $N_{\mu}$, we reconstruct the energy spectra of different mass groups by means of unfolding methods. The procedure and its results, giving evidence for a knee-like structure in the spectrum of iron nuclei, will be presented.}


\begin{document}
\maketitle

\section{Introduction}
The spectrum of cosmic rays follows a power law over many orders of magnitude in energy, overall appearing rather featureless. However, there are a few structures observable. In 1958 Kulikov and Khristiansen \cite{lit:first_knee_measurement} discovered a distinct steepening in the spectrum at around $10^{15}$~eV. Three years later, Peters \cite{lit:peters} concluded that the position of this kink, also called the cosmic ray ``knee'', would depend on the atomic number of the cosmic ray particles if their acceleration and/or propagation is correlated to magnetic fields. Round about half a century later, the KASCADE experiment \cite{lit:kascade_allgemein_nimpaper} clarified that this change in spectral index is caused by a decrease of the so far dominating light\footnote{The description ``light'' refers to the atomic mass of the cosmic ray particles, which are atomic nuclei.} component of cosmic rays \cite{lit:kascade-unfolding}. This result was achieved by means of an unfolding analysis disentangling the manifold convoluted energy spectra of five mass groups from the measured two-dimensional shower size distribution of electrons and muons at observation level. Based on the high energy interaction model QGSJET~01~\cite{lit:qgsjet01}, it was shown that the kink in the all-particle spectrum at around $5\times10^{15}$~eV corresponds to a knee observed in the flux of light primaries. 

Nowadays, there are numerous theories about the origin and acceleration of cosmic rays. Concerning the knee position, some of them predict, in contrast to the rigidity dependence considered by Peters, a correlation with the mass of the particles. Hence, it is of great interest to verify whether also the spectra of the heavy mass \mbox{groups} exhibit analogous structures and if so, at what energies. The KASCADE-Grande experiment \cite{lit:kascade_grande_allgemeim_nimpaper} extends the accessible energy range of KASCADE to higher energies up to $10^{18}$~eV and allows by this to investigate the composition of cosmic rays at regions where the possible, so-called iron knee, is expected. The determination of the iron knee enables the validation of the various theoretical models. Following this purpose, the KASCADE-Grande measurements have been analysed similar to the aforementioned \mbox{studies} \cite{lit:kascade-unfolding} of the KASCADE data. The applied unfolding method will be outlined in the next section. Thereafter, the uncertainties of the analysis and the resulting elemental energy spectra will be shown and studied. A more comprehensive description can be found in \cite{lit:kascade_grande_unfolding,lit:phd}.
\section{Outline of the analysis}\label{sec:outline}
\begin{figure}[!t]
  \centering
  \includegraphics[clip, width=0.8\columnwidth, bb= 0 4 567 360]{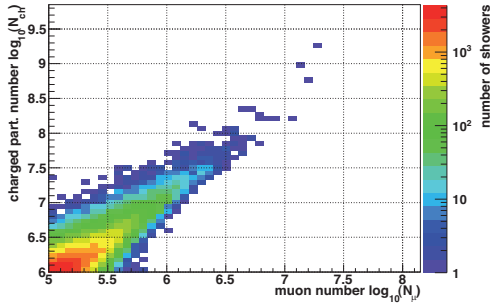}
  \caption{Measured shower size distribution.}
  \label{fig:2d_data_plane}
 \end{figure}

The analysis' objective is to compute the energy spectra of five\footnote{Due to effects of limited resolution, not any number of mass groups can be treated.} cosmic ray mass groups, represented by protons (p) as well as helium (He), carbon (C), silicon (Si), and iron (Fe) nuclei, from $10^{16}$~eV beyond primary energies of $10^{17}$~eV. The two-dimensional shower size spectrum $\mathrm{lg}N_{ch}$ vs. $\mathrm{lg}N_{\mu}$ of charged particles and muons measured with KASCADE-Grande is used as starting point for the unfolding analysis (Fig.~\ref{fig:2d_data_plane}). 
\begin{figure}[!b]
  \vspace{5mm}
  \centering
  \includegraphics[width=0.8\columnwidth]{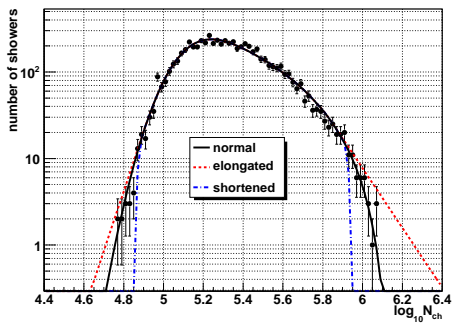}
  \caption{The simulated charged particle number distribution of $2\times10^{15}$~eV proton induced air showers with zenith angles less than $18^{\circ}$. The distribution is fitted based on different approaches (see text).}
  \label{fig:error_est}
 \end{figure}
All measured air showers have to pass certain quality cuts to ensure well reconstructed shower sizes. Particularly, only air showers with zenith angles less than $18^{\circ}$ are used exhibiting at least $10^6$ charged particles and $10^5$ muons. The measurement time covers approximately $1\,318$~days resulting in $78\, 000$ accepted events passing all quality cuts. This corresponds to an exposure of $164\, 709$~m$^2$~sr~yr.

The convolution of the sought-after differential fluxes $\mathrm d J_n/\mathrm d\, \mathrm{lg}E$ of the primary cosmic ray nuclei $n$ into the measured number of showers $N_i$ contributing to the cell $i$ of shower size plane, and thus to the content of this specific charged particle and muon number bin $\left( \mathrm{lg}(N_{ch}), \mathrm{lg}(N_{\mu}) \right)_{i}$ in Fig.~\ref{fig:2d_data_plane}, can be described by an integral equation:
\begin{equation}\label{eq:convolution_integral}
N_i=\sum\limits_{n=1}^{N_\mathrm n} \; \int\limits_{T_{\mathrm m}} \int\limits_{\Omega_{\mathrm{tot}}} \int\limits_{A_{\mathrm f}} \int\limits_{E}  \frac{\mathrm d J_n}{\mathrm d\, \mathrm{lg}E}\; p_n\; \mathrm d\,\mathrm{lg}E\ \cos \theta\ \mathrm dA\ \mathrm d\Omega\ \mathrm dt , \nonumber \\
\end{equation}
with
\begin{eqnarray*}
p_n=p_n\left(   \left(  \mathrm{lg}N_{ch}, \mathrm{lg}N_{\mu}   \right)_i \ | \    \mathrm{lg}E    \right)  \;  .
\end{eqnarray*}
One has to sum over all $N_\mathrm n$ elements contributing to the all-particle cosmic ray spectrum, in this analysis the five representative primaries.  $T_{\mathrm m}$ is the measurement time,  $\Omega_{\mathrm{tot}}$ the total solid angle accessible for the experiment and used for the analysis, and $A_{\mathrm f}$ the chosen fiducial area.  The term $p_n$ represents the conditional probability to reconstruct a certain combination of charged particle and muon number, respectively to get an entry in the cell $\left( \mathrm{lg}(N_{ch}), \mathrm{lg}(N_{\mu}) \right)_{i}$, if the air shower inducing particle was of the type $n$ and had an energy of $E$. More precisely, $p_n$ itself is a convolution combining the intrinsic shower fluctuations occurring whilst the air shower development, the detection and reconstruction efficiency as well as the properties of the reconstruction process of the observables. The cosine term in $\cos \theta\ \mathrm dA$ accomplishes the transformation from the horizontal surface element to the effective detection area.

Equation (\ref{eq:convolution_integral}) can mathematically be understood as a system of coupled integral equations referred to as Fredholm integral equation of first kind. There are various \mbox{methods} to solve such an integral equation, albeit a resolvability often does not \textit{per se} imply uniqueness. In some preliminary tests, it was found that the unfolding algorithm of Gold \cite{lit:gold_alg} yields appropriate and robust solutions. It is an iterative procedure and \textit{de facto} related to a minimization of a chi-square function. For countercheck purposes, all results are validated by means of two additional algorithms, an also iterative method applying Bayes' theorem \cite{lit:bayes_alg} performing very stable, too, and a regularized unfolding based on a combination of the least-squares method with the principle of reduced cross-entropy \cite{lit:entropie_alg}, which yields slightly poorer results.

All these solution strategies have in common that the response\footnote{Also named kernel or transfer function; and, more precisely, it is rather a matrix than a simple function.} function $p_n$ of Eq.(\ref{eq:convolution_integral}) has to be known \textit{a priori}. It is parametrized based on Monte Carlo \mbox{simulations}. The air shower development is simulated by means of CORSIKA~\cite{lit:corsika} 6.307 based on the interaction models \mbox{QGSJET-II-02}~\cite{lit:qgsjet-ii} and \mbox{FLUKA} 2002.4~\cite{lit:fluka}. The experiment's response is simulated using CRES\footnote{\underline{C}osmic \underline{R}ay \underline{E}vent \underline{S}imulation, a program package developed for the KASCADE~\cite{lit:kascade_allgemein_nimpaper} detector simulation.}~1.16/07, which is based on the GEANT~3.21~\cite{lit:GEANT3_21} detector description and simulation tool. 
\section{Error analysis}
The determination of the elemental energy spectra will be subjected to influences of different error sources. They can roughly be classified in two categories: uncertainties induced, or at least appearing whilst the deconvolution process, as well as those embedded in the computed response function caused by the limited Monte Carlo statistics and by the uncertainties of the interaction models used. 
\begin{figure*}[!t]
   \centerline{\includegraphics[width=0.9\columnwidth]{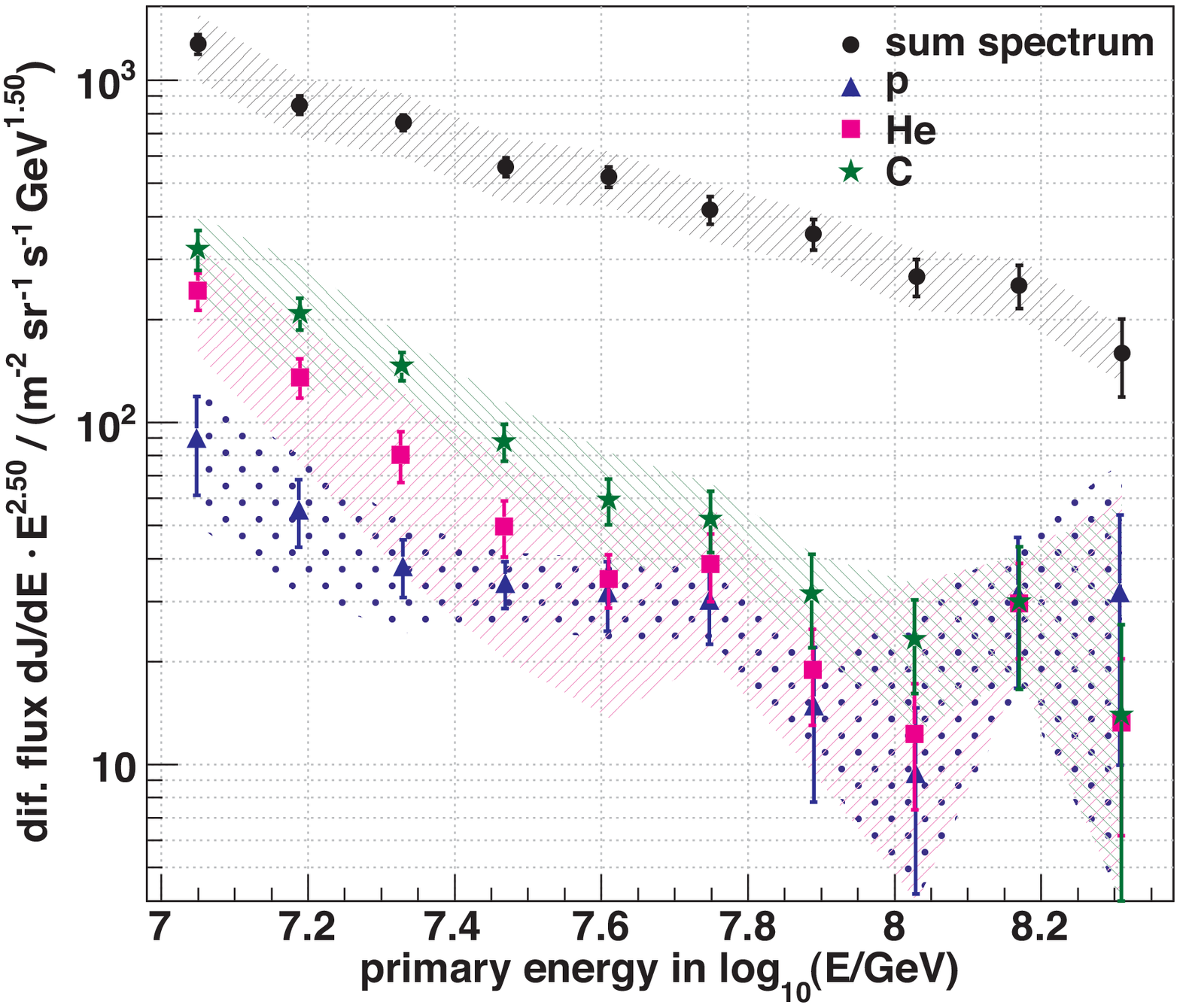}\label{fig2}
              \includegraphics[width=0.9\columnwidth]{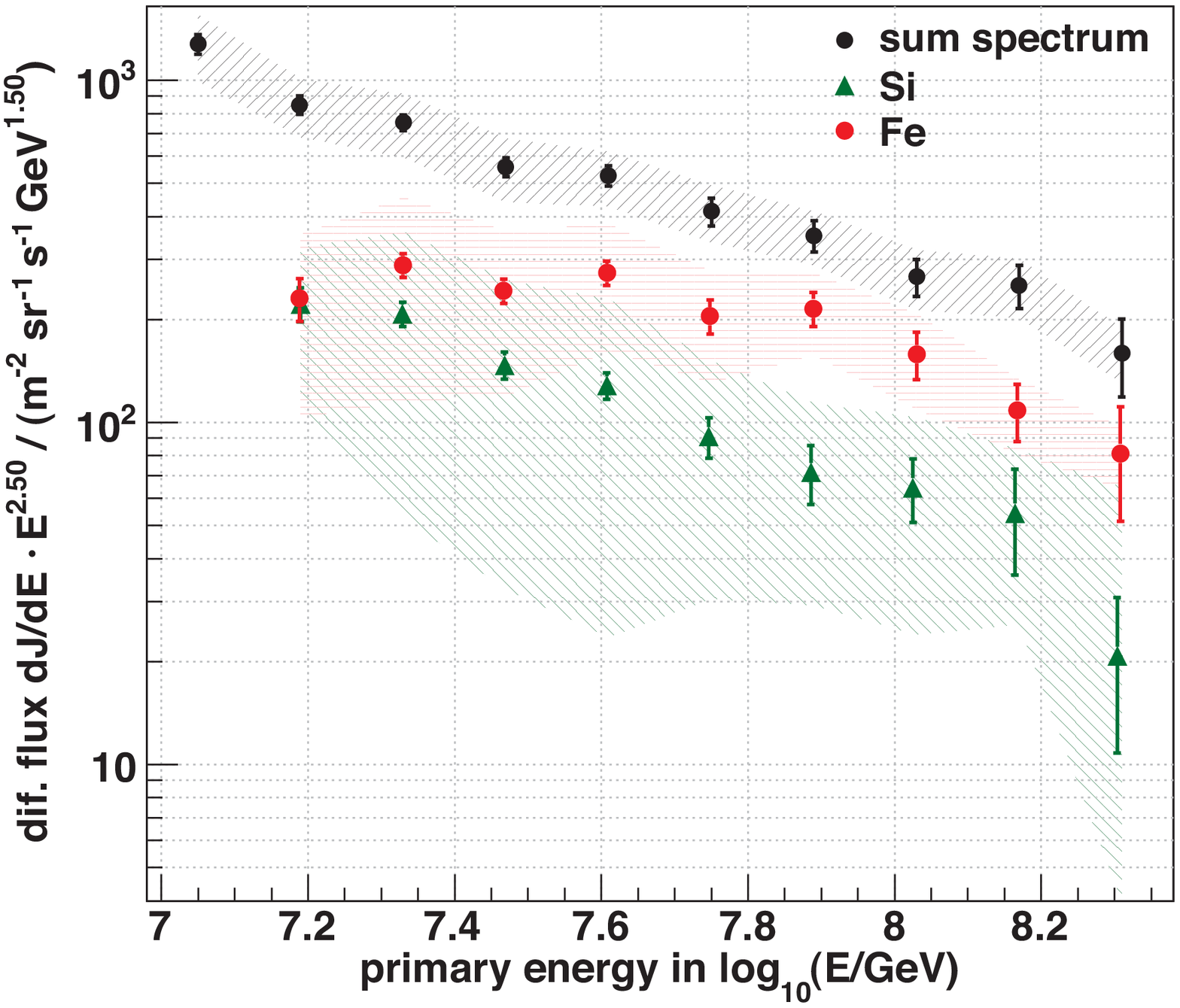} \label{fig3}
             }
   \caption{The unfolded energy spectra for elemental groups of cosmic rays, represented by protons, helium, and carbon nuclei (left panel) as well as by silicon and iron nuclei (right panel), based on KASCADE-Grande measurements. The all-particle spectrum, which is the sum of all five individual spectra, is also shown. The error bars represent the statistical uncertainties, while the error bands mark the maximal range of systematic uncertainties (see text). The response matrix used is based on the interaction models \mbox{QGSJET-II-02}~\cite{lit:qgsjet-ii} and \mbox{FLUKA} 2002.4~\cite{lit:fluka}.}
   \label{fig:unfolded_spectra}
 \end{figure*}
\subsection{Uncertainties whilst the deconvolution}
Firstly, the used data set is only a small sample based on a limited exposure, and hence suffering from statistical uncertainties. They are propagated through the unfolding algorithm and affect the quality of the solution. Furthermore, the used deconvolution method itself can introduce a systematic bias. The influences of both sources can be evaluated by means of a frequentist approach. Assuming appropriate
spectral indices, some trial elemental energy spectra are specified based on which a test data sample can be generated using Eq.(\ref{eq:convolution_integral}). Subsequently, these data samples are unfolded. Since the true solution is \textit{a priori} known, the deconvolution result can be compared to it to reveal statistical fluctuations induced by the limited measurement time and a possible systematic bias induced by the unfolding method.
\subsection{Influences of limited Monte Carlo statistics} 
The amount of simulated air showers is strongly limited due to computing time. Due to the limited Monte Carlo statistics, the computation of the response function, i.e. the parametrization of the intrinsic shower fluctuations as well as of the detector properties, will only be possible under certain uncertainties resulting in a systematic error of the finally unfolded solution. In Fig.~\ref{fig:error_est}, the simulated charged particle number distribution in case of proton induced air showers with primary energy of $2\times10^{15}$~eV is shown exemplarily. A scattering around the used parametrization (``normal'') can be observed. This statistical uncertainty will be treated conservatively: Considering the computed fit parameters and their errors, some new sets of parameters are calculated by means of a random generator. Based on each set, new response functions can be computed and used to unfold the data. Comparing the results reveals the caused systematic uncertainty in the solution.

The distributions' tails have to be inspected in more detail. Because of the very low statistics, the tails can vary within a certain range without worsen the fit result. In particular, the right tail describing the fluctuations in direction to higher energies can have an important impact on the unfolded solution due to the steeply falling flux of cosmic rays. The systematic influence of the tails will be estimated conservatively by computing two additional response functions assuming in contrast to the standard case either a very fast decreasing or an elongated tail (cf. Fig.~\ref{fig:error_est}). Using both for a deconvolution and comparing the results yields the maximal systematic error range caused by the uncertainty in the tails description. 
\subsection{Uncertainties of interaction models used}
D'Enterria et al. \cite{lit:lhc_models} compared the first Large Hadron Collider (LHC) data with the predictions of various Monte Carlo event generators, including the model \mbox{QGSJET-II} used in this analysis. They stated that none of the investigated models can describe consistently all measured observables at the LHC, but, that there is basically a reasonable overall agreement. Nevertheless, it was shown in \cite{lit:kascade_grande_unfolding,lit:phd} based on this unfolding analysis that the model QGSJET-II-02 yields results, which agree with the data measured with KASCADE-Grande. The uncertainties caused by the models used are difficult to quantify as \mbox{all} models can fail if new physics is appearing in this energy range. However, in \cite{lit:kascade-grande-all_particle,lit:kascade-unfolding,lit:KG_influence_hadr_int} it was shown that the high energy interaction model affects primarily the relative abundances of the mass groups and the absolute scale in energy assignment, while specific structures in the spectra are conserved. In addition, it is known that the low energy interaction model has less influence on the final result, as already the analyses based on the KASCADE measurements have proved \cite{lit:KASCADE_low_energy_interaction_model}. 

\section{Results and conclusion}
In Fig.~\ref{fig:unfolded_spectra}, the unfolded differential energy spectra of lighter primaries (protons as well as helium and carbon nuclei, left panel), and the spectra of heavier ones (silicon and iron nuclei, right panel) are depicted. In addition, all five unfolded spectra are summed up to the all-particle flux. The shaded band indicates the methodical uncertainties, while the error bars represent the statistical error originating from the limited measurement time. 

With increasing energy, the heavy component becomes the dominant contributor to the cosmic ray composition. This agrees with the results of KASCADE \cite{lit:kascade-unfolding}, where a reduction of the light component beyond the first knee was found.

The spectra of lighter primaries are rather featureless within the given uncertainties. However, there are slight indications for a recovery of the proton spectrum at higher energies, which is not significant, however. Though, the recovery would agree with the significant ankle-like feature in the energy spectrum of light elements reported in \cite{lit:kascade-grande-proton-recovery}, where another analysis method was applied to the KASCADE-Grande data.

In the iron spectrum, there is a slight bending discernible at around $1\times 10^{17}$~eV. The position of this knee-like structure agrees with the one in the all-particle spectrum at around $1\times 10^{17}$~eV reported in \cite{lit:kascade-grande-all_particle}, such that both features seem to be correlated. Furthermore, the position of the knee in the iron spectrum is compatible with the one of a significant kink in the spectrum of heavy primaries observed in \cite{lit:kascade-grande-heavy-knee}.

\begin{figure}[t]
 \vspace{2mm}
  \centering
  \includegraphics[clip, width=0.9\columnwidth, bb=0 0 567 368]{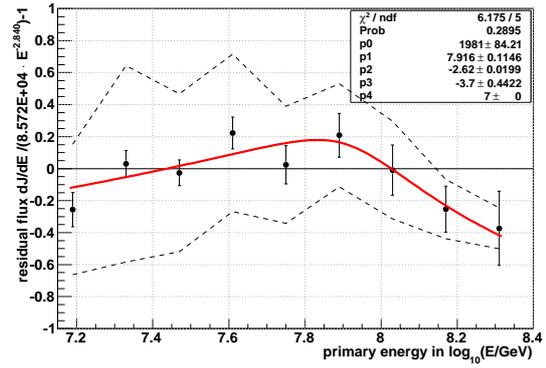}
  \caption{Residual flux between the iron spectrum shown in Fig.~\ref{fig:unfolded_spectra} and a spectrum that was derived by a single power law fit to that iron spectrum. Additionally, the iron spectrum is now fitted by a double power law.}
  \label{fig:residu}
 \end{figure}

In order to judge the structures in the unfolded iron spectrum, it is fitted preliminarily by a single power law. However, the resulting chi-square probability for such a featureless single power law is below 1\% ($\chi^2/\mathrm{\textit{ndf}}=18.9/7$). In Fig.~\ref{fig:residu}, the residual flux between the iron spectrum shown in Fig.~\ref{fig:unfolded_spectra}, and such a spectrum that is derived by a single power law fit is depicted. Additionally, the iron spectrum is fitted by a double power law:
\begin{equation}\label{eq:knee}
\frac{\mathrm d J(E)}{\mathrm d\, \mathrm{lg}E}=
p_0 \times E^{p_2} \times \left( 
1+
\left( 
\frac{E}{p_1}
\right) ^{p_4}
\right)^{\left(  p_3-p_2  \right)/p_4}\;  ,
\end{equation}
where $p_1=\mathrm{lg}(E_{\mathrm{knee}}/\mathrm{GeV})=7.9\pm 0.1$ is the knee position, while $p_2=-2.62\pm0.02$ and $p_3=-3.7\pm0.4$ are the spectral indices below and above the knee. The sharpness of the knee structure is encoded in $p_4=7.0$ and was fixed without worsening the quality of the fit, while $p_0$ is a free normalization parameter. This fit describes the spectrum significantly better (chi-square probability at around 30\% with $\chi^2/\mathrm{\textit{ndf}}=6.2/5$), giving strong indications for a true kink in the iron flux at around 80~PeV. 

Comparing\footnote{And assuming that the mass groups represented by p and Fe actually consist only, or at least primarily, of those two primaries.} the position of this potential iron knee to that for protons (at around 3~PeV to 5~PeV, cf. \cite{lit:kascade_grande_unfolding,lit:phd}) gives indications for a scaling of the knee positions with the charge of the nuclei, encouraging the cosmic ray acceleration models based on magnetic fields. 

To summarize, this analysis gives strong indications for a kink in the iron-like cosmic ray spectrum at around 80~PeV, as well as for a dependence of the cosmic ray acceleration process on the charge of the nuclei, both on the premise that
especially the used model QGSJET-II-02 describes the physics of hadronic interactions with a high level of reliability at these energies. 

\vspace*{0.5cm}
\footnotesize{{\bf Acknowledgment:}{The authors would like to thank the members of the engineering and technical staff of the KASCADE-Grande collaboration, who contributed to the success of the experiment. The KASCADE-Grande experiment is supported in Germany by the BMBF and by the Helmholtz Alliance for Astroparticle Physics - HAP funded by the Initiative and Networking Fund of the Helmholtz Association, by the MIUR and INAF of Italy, the Polish Ministry of Science and Higher Education, and the Romanian Authority for Scientific Research UEFISCDI (PNII-IDEI grants 271/2011 and 17/2011).}}

\end{document}